\documentstyle[ 12pt]{article}

\begin{document}

\centerline { \bf  ON MAXIMAL-ACCELERATION , STRINGS AND THE  GROUP }
\centerline{ \bf OF MINIMAL PLANCK-AREA RELATIVITY THEORY }

\bigskip

\centerline { Carlos Castro }

\centerline{ Center for Theoretical Studies of Physical Systems, }

\centerline{ Clark Atlanta University, Atlanta, GA. 30314 }

\centerline { November  , 2002 }

\bigskip

\centerline{\bf  Abstract}

\bigskip

Recently we have presented a new physical  model that links the maximum
speed of light with the minimal Planck scale into a 
maximal-acceleration
Relativity principle in phase spaces .  The maximal proper-acceleration
bound is  $a = c^2/ \Lambda$ where $ \Lambda$ is the Planck scale.
The group transformation laws of this Maximal-acceleration Relativity 
theory
under velocity and 
acceleration boosts are analyzed in full detail.  For pure acceleration
boosts it is shown that the
minimal Planck-areas ( maximal tension ) are universal invariant 
quantities
in any frame of reference.
The implications of this minimal Planck-area ( maximal tension ) 
principle
in future developments of string theory,  $ W$-geometry and Quantum 
Gravity
are briefly outlined.

\bigskip

\centerline {\bf I .  Introduction }

\bigskip

In recent years there has been growing evidence  that the Relativity
principle should be extended to
include all dimensions and signatures on the same footing .
Relativity in C-spaces (Clifford manifolds) [1] is a very natural 
extension
of Einstein's relativity and Nottale's scale relativity [2] where the
impassible speed of light and the minimum Planck scale are the two 
universal
invariants.  An event in C-space is represented by a   $polyvector$ , 
or
Clifford-aggregate of lines, areas, volumes, ...... which  bear a 
one-to-one
correspondence to the holographic shadows/projections (onto the 
embedding
spacetime coordinate planes) of a nested family of $p$-loops (closed 
$p$-
branes of spherical topology) of various dimensionalities: $p = 0$
represents
a point; $p = 1$ a closed string, $p = 2$ a closed membrane, etc.... 
where
$ p = 0, 1, 2, ....  D-1$.

The invariant ``line"  element associated with a polyparticle is:

$$ d\Sigma ^2 = dX.dX =  ( d \Omega) ^2 +  \Lambda ^{2D-2} ( 
dx_{\mu}dx^
{\mu} ) + 
\Lambda ^{ 2D -4 } ( d x_{\mu\nu} ) ( d x^{\mu\nu} ) + ...\eqno(1.1) $$
the Planck scale appears as a natural quantity in order to match units 
and
combine p-branes ( p-loops ) of different dimensions.   The 
polyparticle
lives in a target background of  $D + 1 = p+2 $  dimensions due to the 
fact
that C-space has $two$ times , the coordinate time $ x_o = t $ and the 
$
\Omega$ temporal variable representing the proper $ p+1$-volume.  The 
fact
that the Planck scale is a minimum
was based on the real-valued interval $ dX $ when $ dX.dX > 0 $. The 
analog
of photons in C-space are $tensionless$ branes : $ dX.dX = 0 $. Scales
smaller than 
$\Lambda$ yield " tachyonic " intervals $ dX. dX < 0 $ [1]. Due
to the matrix representation of the gamma matrices and the cyclic trace
property, it can be easily seen why the line element is invariant under 
the
C-
space Lorentz $group$  transformations:
$$Trace ~ X'^{2 } = Trace~ [ R X^2 R^{ - 1 } ] = Trace~ [ R R^{ -1 } 
X^2 ] =
Trace~X^2 ~, \eqno( 1.2) $$
where a finite polydimensional rotation that reshuffles dimensions is
characterized by the C-space ``rotation" matrix:

$$  R = \exp [ i (\theta I + \theta ^{\mu} \gamma _{\mu} + 
\theta^{\mu\nu}
\gamma _{\mu \nu} + ...)] .  \eqno(1.3) $$
The parameters $\theta, \theta ^{\mu}, \theta ^{\mu\nu}, ......$ are 
the
C-space
extension of the Lorentz boost parameters and for this reason the naïve
Lorentz transformations of spacetime are modified to be:

$$ x'^{\mu} = L^{\mu}_{\nu} ~ [ \theta , \theta ^\mu, \theta^{\mu\nu} , 
...]
x^{\nu}~ +  L^\mu_{\nu \rho } ~ [ \theta, \theta^\mu , \theta^{\mu\nu} 
,
...] x^{\nu \rho } + .....  \eqno (1.4 ) $$

It was argued in [1] that the extended Relativity principle in C-space 
may
contain the clues to unravel the physical foundations of string and 
M-theory
since the dynamics in C-spaces encompass in one stroke the dynamics of 
all
p-
branes of various dimensionalities. In particular, how to formulate a 
master
action that encodes the collective dynamics of all extended objects.

For further details about these issues we refer to  [1 ] and all the
references therein. Like the derivation of the minimal length/time
string/brane uncertainty relations;  the logarithmic corrections to the
black-
hole area-entropy relation;  the existence of a maximal Planck 
temperature ;
the origins of a higher derivative gravity with
torsion;  why quantum-spacetime may be  truly infinite dimensional 
whose
average dimension today is close to $ 4 + \phi^3  = 4.236...$ where $ 
\phi
= 0.618...$ is the Golden Mean ;
the construction of the p-brane propagator;  the role of
supersymmetry;  the emergence of two times; the reason behind a running
value for $ \hbar $ ;   the way to correctly pose the cosmological 
constant
problem as well as other results.

In  [ 1  ]  we discussed another physical model that links the maximum
speed of light , and the minimal Planck scale,  into a 
maximal-acceleration
principle in the spacetime tangent bundle, and consequently, in the 
phase
space (cotangent bundle). Crucial in order to establish this link was 
the
use
of Clifford  algebras in phase spaces. The maximal proper acceleration 
bound
is  $a = c^2/ \Lambda$  in full agreement with [4] and the Finslerian
geometry point of view in [6].  A  series of reasons why  C-space 
Relativity
is more physically appealing
than all the others proposals based on kappa-deformed Poincare algebras 
[ 10
] and other quantum algebras was presented .  Maximal-acceleration 
effects
within the context of kappa-deformed Poincare have been discusssed in [ 
11
].

On the other hand, we argued why the truly $bicovariant$ quantum 
algebras
based on inhomogeneous quantum groups developed by Castellani  [ 15  ]  
had
a very interesting feature related to the $T$-duality in string theory 
;
the deformation $ q$ parameter could be written as :  $ q = exp [ 
\Lambda/ L
] $  and, consequently, the classical limit  $ q = 1 $ is attained when 
the
Planck scale $ \Lambda$ is set to zero, but also when the upper 
impassible
scale $ L $ goes to infinity !  This entails that there could be $two$ 
dual
quantum gravitational theories with the same classical limit !  Nottale 
has
also postulated that if there is a minimum Planck scale, by duality, 
there
should be another upper impassaible upper scale $ L $ in Nature [ 2 ] .
For a recent discussion on maximal-acceleration and kappa-deformed 
Poincare
algebras see  [ 10 ] .  It was  also argued in [ 1 ]  why the theories 
based
on kappa-deformed Poincare algebras may in fact be related to a Moyal
star-product deformation of a classical Lorentz algebra whose 
deformation
parameter is precisely the Planck scale $\Lambda = 1/ \kappa$.

In section {\bf 2.1 },  we  review once again  the work in [ 1 ]  and 
show
how to derive the Nesterenko action [5] associated with a sub-maximally
accelerated particle in spacetime directly from phase-space Clifford
algebras and present a  full-fledged C-phase-space generalization of 
the
Nesterenko action .  It was this principle of Maximal-Acceleration 
Phase
space Relativity that allowed us
to derive the exact integral equation that governs the
Renormalization-Group-like scaling dependence of the fractional change 
of
the fine structure constant as a function of the cosmological redshift
factor and a cutoff scale $ L_c $ where the maximal acceleration
relativistic effects are dominant.
For a review of the variation of the fundamental constants see Uzan [ 
13 ] .
Maximal-acceleration corrections to the Lamb shifts of one-electron 
atoms
were performed by 
Lambiase, Papini and Scarpetta [ 13 ].

If the cutoff scale $ L_c$ was set equal to the minimal Planck scale we 
have
shown [ 1 ] why one obtains a Cosmological model  dominated $entirely$ 
by
the cosmological constant,  with $ \Omega_\Lambda  =   1 $.
We argued why in this extreme case scenario all the matter in the 
Universe
may have been created
out of the vacuum  ( vacuum fluctuations ) as a result of the 
acceleration
effects , analogous to the Hawking-Unruh effect of particle production 
due
to  the accelerated motion ( noninertial ) with respect to a Minkowski
vacuum.

In section { \bf 2.2  }  we review the main features of Born`s Dual
Relativity and the work of Low [ 14 ]  in the construction of $ U ( 1, 
3 ) $
group transformations which leave the intervals in classical Phase 
spaces
invariant along with the construction of the unitary irreducible
respresentations which describe the particle spectrum of the theory.   
In
section {\bf 3 } we present the explicit transformations rules of the 
Phase
space coordinates under velocity and acceleration boosts and show 
explicitly
why they have the required  $group$  structure to qualify for a  
genuine
Phase Space Relativity Theory.

In section {\bf 4 } we  prove why pure acceleration-boost 
transformations
leave $invariant$ the  $minimal$  Planck-Areas  ,  fact
which can be reinterpreted as  the existence of an invariant universal
maximum string tension  and Planck temperature in Nature.

We finalize by making some comments concerning the Conformal Group,
$ W$ gravity ,  higher conformal spin theories on $ AdS$ spaces ,  $
W_\infty$ strings .....
and advocate the importance to build an Extended  Relativity Theory in
C-Phase-Spaces 
that will encompass the physics of $all$  p-branes  into  one single 
footing
by 
implementing the Relativity principle of $minimal$  and $invariant$
Planck areas ,  Planck-volumes,  Planck-hypervolumes ...in all frames 
of
references in Phase spaces .

  \bigskip

\centerline{ \bf 2. Maximal-Acceleration Phase Space Relativity }
\bigskip

\centerline { \bf  2.1 .  Maximal-Acceleration from Clifford algebras }

  \bigskip

Redaers familiar with the previous work may omit this subsection.
We will follow closely the procedure described in the book [3] to 
construct
the phase space Clifford algebra. For simplicity we shall begin with a 
two--
dimensional phase space, with one coordinate and one momentum variable 
and
afterwards we will generalize the construction  to higher dimensions.

Let $ e_{p}   e_{q} $ be the Clifford basis elements in a 
two--dimensional
phase space obeying the following relations:

$$  e_p . e_q \equiv  { 1 \over 2 } ( e_q e_p + e_p e_q ) = 0. ~~~ e_p 
. e_p
= e_q . e_q = 1 . \eqno(2.1) $$

The Clifford product of  $ e_p, e_q $ is by definition the sum of the 
scalar
product and wedge product furnishing the unit $bivector$:

  $$ e_p e_q \equiv e_p . e_q +  e_p \wedge e_q  = e_p \wedge e_q =  j  
.
~~~
j^ 2 =  e_p e_q  e_p e_q = - 1 . \eqno (2.2 ) $$
due to the fact that $ e_p, e_q $ anticommute, eq.~( 2.1).

In this fashion, using Clifford algebras one can justify the origins of
complex numbers without introducing them ad-hoc. The imaginary unit $ j 
$ is
$ e_p e_q $.  For example, a Clifford vector in phase space can be 
expanded,
setting aside for the moment the issue of units, as:

$$  Q = q e_q  + p e_p . ~~~ Q e_q = q + p e_p e_ q = q + j p = z. ~~~
e_q Q = q + p e_q e_p = q - j p = z^*~, \eqno ( 2.3 ) $$
which reminds us of the creation/annhilation operators used in the 
harmonic
oscillator case and in coherent states.

The analog of the action for a massive particle is obtained by taking 
the
scalar product: 
$$ dQ . dQ = (dq)^2 + ( dp)^2  \Rightarrow S = m \int  \sqrt { dQ. dQ } 
=
m \int  \sqrt {( dq)^2 + (dp)^2 } . \eqno ( 2.4 ) $$

One may insert now the appropriate length and mass parameters in order 
to
have consistent units:

$$ S = m \int  \sqrt {  ( dq )^2 + ( {  \Lambda \over m  } )^2  (dp)^2 
}.
\eqno (2.5 ) $$
where we have introduced  the Planck scale $\Lambda$  and the mass $m$ 
of
the
particle to have consistent units, $\hbar = c = 1$.
The reason will become clear below.

Extending this two-dimensional action to a higher $2n$-dimensional 
phase
space requires to have
$ e_{p_{\mu }} ,  e_{q_{\mu}} $ for the Clifford basis where $\mu = 1, 
2,
3...n$. The action in this $2n$-dimensional phase space is:
$$ S = m \int  \sqrt {  ( dq^{\mu} dq_{\mu} ) + ({\Lambda \over m  } 
)^2
(dp^ {\mu} dp_{\mu} ) } =
m \int  d \tau \sqrt { 1  + ( {  \Lambda \over m  } )^2  (dp^{\mu}/ d 
\tau )
(  dp_{\mu}/ d \tau  ) }
\eqno (2.6 ) $$
in units of $c = 1$, one has the usual infinitesimal proper time
displacement  $d \tau ^2 = dq^{\mu}  dq_{\mu}$.

One can easily recognize that this action (2.6), up to a numerical 
factor of
$ m/a$,  is nothing but the action for a sub-maximally accelerated 
particle
given by
Nesterenko  [5].  It is sufficient to rewrite: $dp^\mu / d \tau  = m 
d^2
x^\mu / d \tau ^2 $  to get from eq.~(2.6):
$$ S = m \int  d \tau \sqrt {  1  + \Lambda^2  (d^2 x^ \mu/ d \tau^2  ) 
(
d^2 x_\mu/ d \tau^2) } . \eqno ( 2.7) $$

Using the postulate that the maximal-proper acceleration is given in a
consistent manner with the minimal length principle (in units of $c = 
1$):

$$ a = c^2 / \Lambda  = 1/\Lambda \Rightarrow
S = m \int  d \tau \sqrt {  1  +   (  { 1 \over a }  )^2  (d^2 x^ \mu/ 
d
\tau^2  ) (  d^2 x_\mu/ d \tau^2   ) }. \eqno ( 2.8) $$
which is exactly the action of [5], up to a numerical factor of $ m/a 
$,
when  the metric signature is $ ( +, -, - , - ) $.

The proper acceleration is $orthogonal$ to the proper velocity as a 
result
of
differentiating the timelike proper velocity  squared:

$$ V^2 = {dx^{\mu} \over d \tau } { d x_{\mu} \over d \tau } = 1 =   
V^\mu
V_\mu > 0 \Rightarrow  { d V^{\mu} \over d \tau } V_\mu =  { d^2 
x^{\mu}
\over d \tau ^2 }
V_\mu =  0~, \eqno(2.9)  $$
which means that if the proper velocity is timelike the proper 
acceleration
is spacelike so that:

  $$  g^2 ( \tau ) \equiv -  (d^2 x^ \mu/ d \tau^2  ) (  d^2 x_\mu/ d
\tau^2)
 > 0  \Rightarrow
S = m \int  d \tau \sqrt {  1  -    { g^2  \over a^2  }     } \equiv  m 
\int
d \omega ~, \eqno(2.10 ) $$
where we have defined:

$$  d \omega  \equiv \sqrt {  1  -    { g^2  \over a^2  }     } d \tau 
.
\eqno
(2.11) $$
The dynamics of a submaximally accelerated particle in Minkowski 
spacetime
can be reinterpreted as that of a particle moving in the spacetime 
$tangent-
bundle$ background whose $Finslerian$-like metric is:

$$  d\omega^2 = g_{\mu \nu} ( x^\mu, dx^\mu ) dx^\mu dx^\nu  =
(d \tau)^2  (  1  -    { g^2  \over a^2  }     )                .
\eqno(2.12)
$$

For uniformly accelerated motion, $ g ( \tau ) = g = constant$ the 
factor:

$$ { 1 \over \sqrt {  1  -  { g^2  \over a^2  }  }   }  \eqno (2.13) $$
plays a similar role as the standard Lorentz time dilation factor in
Minkowski spacetime.

The action is real valued if, and only if, $ g^2  < a^2 $ in the same 
way
that the action in Minkowski spacetime is real valued if, and only if, 
$v^2
<
c^2$. This explains why the particle dynamics has a bound on proper-
accelerations. Hence, for the particular case of a $uniformly$ 
accelerated
particle whose trajectory in Minkowski spacetime is a hyperbola, one 
has an
Extended  Relativity of $uniformly$ accelerated observers whose proper-
acceleration have an upper bound given by  $c^2/ \Lambda$. Rigorously
speaking, the spacetime trajectory is obtained by a canonical 
projection of
the spacetime tangent bundle onto spacetime. The invariant time, under 
the
pseudo-complex extension of the Lorentz group [8], measured in the 
spacetime
tangent bundle is no longer the same as $\tau$,  but instead, it is 
given by
$\omega ( \tau )$.

This is similar to what happens in C-spaces, the truly invariant 
evolution
parameter is not
$\tau$ nor $\Omega$, the Stuckelberg parameter [3], but it is $\Sigma$
which
is the world interval in C-space and that has units of $length ^ D$. 
The
$group$ of C-space Lorentz transformations preserve the norms of the
Polyvectors and these have units of hypervolumes; hence C-space Lorentz
transformations are volume-preserving.

Another approach to obtain the action for a sub-maximally accelerated
particle was given by [8] based on a pseudo-complexification of 
Minkowski
spacetime and the Lorentz group that describes the physics of the 
spacetime
tangent bundle. This picture is not very different form the Finslerian
spacetime tangent bundle point of view of Brandt [6].  The invariant 
group
is
given by a pseudo-complex extension of the Lorentz group acting on the
extended coordinates $ X = a x^\mu + I v^\mu $ with $ I^2 =  1 $ 
(pseudo-
imaginary unit) where both position and velocities are unified on equal
footing. The invariant line interval is
$a^2 d \omega^2 = (dX)^2$.

A C-phase-space generalization of these actions (for sub-maximally
accelerated particles, maximum tidal forces) follows very naturally by 
using
polyvectors:
$$ Y =  q^\mu e_{ q_\mu  } +   q^ { \mu \nu } e_{ q_\mu  } \wedge e_{
q_\nu  } +
q^ { \mu \nu\rho } e_{ q_\mu  } \wedge e_{ q_\nu  } \wedge e_{q_\rho} +
....$$
$$ + p^\mu e_{ p_\mu  } +   p^ { \mu \nu } e_{ p_\mu  } \wedge e_{ 
p_\nu  }
+ ...~,\eqno (2.14) $$
where one has to insert suitable powers of $\Lambda$ and $m$ in the
expansion
to match units.

The C-phase-space action reads then:

$$S \sim \int \sqrt { dY . dY } = \int \sqrt {   dq^\mu dq_\mu +  d 
q^{\mu
\nu } dq_{\mu \nu } + ... + dp^\mu dp_\mu +  d p^{\mu \nu } dp_{\mu \nu 
}
+ .....}~.
\eqno (2.15 ) $$

This action is the C-phase-space extension of the action of Nesterenko  
and
involves quadratic derivatives in C-spaces which from the spacetime
perspective are effective {\em higher} derivatives theories [1 ]  where 
it
was shown why the scalar curvature in C-spaces is equivalent to a  
higher
derivative gravity. One should expect a similar behaviour for the 
extrinsic
curvature of a polyparticle motion in C-spaces. This would be the 
C-space
version of the action for rigid particles [7]. Higher derivatives are 
the
hallmark of {\em W}-geometry (higher conformal spins).

Born-Infeld models  have been connected to  maximal-acceleration [8]. 
Such
models admits an straightforwad formulation using the geometric 
calculus of
Clifford algebras. In particular one can rewrite all the nonlinear 
equations
of motion in precise Clifford form [9]. This lead that author to 
propose the
$nonlinear$ extension of Dirac's equation for massless particles due to 
the
fact that spinors are nothing but right/left ideals of the Clifford 
algebra:
i.e., columns, for example, of the  Maxwell-Field strength bivector
$F = F_{\mu\nu} \gamma ^{\mu} \wedge \gamma ^{\nu}$.

\bigskip

\centerline{ \bf 2.2 Born's Dual Relativity Principle }
\smallskip

Long time ago Max Born [ 14 ] proposed the reciprocally-conjugate 
Relativity
principle that states
that physics in Phase spaces must be invariant under the reciprocity
transform :

$$  \{ Q ,  P \}  \rightarrow  \{  P,  - Q  \}  . \eqno ( 2.16 ) $$

such that the rates of change of momentum ( force ) is bounded by a
universal constant 
$ b $ .  In units of $ \hbar = c = 1 $ we have that the 
maximal-acceleration
is given in terms of the Planck
scale $  A =  ( 1/ L_P) $.  Hence the maximal force  subjected by an
elementary particle =
Planck-mass x acceleration = $ ( 1 / L_P^2 ) $  coincides with the 
maximal
string tension in this sytem of units.

The appropriate group of dynamical symmetries that incorporates 
elementary
particle states has been studied recently by  Low [ 14 ]  in terms of 
the
canonical group $ C ( 1, 3 ) $ ,
acting in an extended noncommuting Phase space which  is given by the
$semidirect$ product of $ U ( 1, 3 ) = SU(1, 3 ) \times U(1) $ with the
Weyl-Heisenberg  group $ H ( 1, 3 ) $.   An essential point  that
distinguishes Born`s Dual Relativity from others is that here the 
Canonical
group $ C ( 1, 3 ) $ is traded for the naive Poincare group ! In 
ordinary
Relativity,  Minkowski space is represented as the coset space of the
Poincare group modulo de Lorentz group.
In this dual Relativity  theory the physical $Noncommuting$  Phase 
space  is
comprised of  ' ' quantum oscilators ' ',  instead of   ' ' points ' ',
and 
is represented by the coset space  $ C ( 1, 3 )/ SU(1, 3 ) $ whose 
metric is
the 
invariant  second Casimir of the canonical group :

$$ T^2 + { E^2 \over c^2 b^2 } - X^2 - { P^2 \over b^2 }  +
{ 2 \hbar I \over b c } (  { Y \over bc } - 2 ) . \eqno ( 2.17) $$

The generators of the canonical algebra are :

$$ T, E,  Q^i , P^i , I , Y   . \eqno ( 2.18 ) $$
 where $ I $ is the center of the Weyl-Heisenberg algebra and $ Y $ a $ 
U (
1 ) $ generator.  
Hence , an element of the unitary group $ U ( 1, 3 ) $ may be 
represented
as : 

$$  e^Z. ~~~ where ~  Z =  { \xi_v^i \over c } K_i + { \xi_a^i \over b 
} N_i
+ \alpha^i J_i +  \theta^{ ij } M_{ij} +
{\phi \over bc } Y . \eqno ( 2.19 ) $$
where $ K_i $ is the velocity-boost  generator ; $ N_i $ is the
acceleration-boost generator ;  $ J_i $ is  the generator of spatial
rotations in the $ x_j-x_k$ plane.  $ M_{ij} $ is the generator of
rotations in the $ p_i-p_j$ direction.  $ Y $ is the generator of $ U ( 
1 )
$ . 

The Heisenberg group is represented  as :

$$ e^A. ~~~where ~  A = \sqrt {  { bc \over  \hbar }  } ~
( t T + { e \over bc } E + { q^i \over c} Q_i  + { p^i \over b } P_i + 
{
\eta \hbar \over bc } I ) .
\eqno ( 2.20 ) $$ 

For more details we refer to Low [ 14 ] .  One of the most important 
results
in [ 14 ] was the construction of unitary irreducible respresentations 
(
unireps ) of the canonical group using Mackey's theory of induced
representations.  These unireps contain representations of $ U ( 1, 3 ) 
$
which was proposed long ago by Kalman [ 14 ] as a dynamical group for
hadrons.  String theory originated with the study of hadrons, hence it 
is
not surprising to have come back to the initial starting point.  The 
unireps
contain discrete series representations that can be decomposed into
$infinite$ ladders where the rungs are representations of
$ U ( 3 ) $. 

The most salient feature of these representations that $define$ the
$particle$  states is that an acceleration-boost will transform a
single-particle state into a $composite$ state,  which in turn,  can be
decomposed into a sum of single particle states representing the 
particle
interactions in  the accelerated-frame of reference. This is yet 
another
realization of the Hawking-Unruh effect of particle creation when an
observer is moving in an accelerated frame of reference with respect to 
a
Minkowski vacuum.  
A thermal radiation of particles is detected whose temperature, 
measured
with respect to an asymptotic observer in the case of particle emission 
by
Black Holes , is  linearly proportional to the acceleration
$ T = { \hbar a \over 2 \pi k_B c } $ .  Where $ k_B$ is Boltzman 
constant.
The maximal-acceleration Phase space Relativity principle  requires a
maximum Planck Temperature  of the order of $ 10^{ 32 } $ Kelvin.
   
Most importantly, for the present author,  is the  phenomenon that a 
single
particle state can be boosted via acceleration-boosts into a composite
multiple-particle state and which is yet another manifestation of the
Bootstrap principle of String theory ;  i.e. $all $ particles are made 
of
each other.  
In addition, the rest and null frames automatically yielded  the groups 
$ SU
( 3) , SU(2 ) , U(1 ) $ that appear in the standard theory of the 
strong,
weak and electromagnetic interactions $without$  the need to compactify 
from
higher dimensions  to four dimensions.

In the next section we will study the Group properties of the
Maximal-accerelation Phase space relativity
in the $commuting$ Phase space case.

\centerline{\bf  3 .  Maximal-Acceleration Relativity and the $ U(1,3 
)$
Group transformations   }

\bigskip

The $ U ( 1, 3 ) = SU(1, 3 ) \otimes U(1 ) $  Group transformations  [ 
14 ]
can be simplified drastically when the velocity/acceleration boosts are
taken to lie in the
$z = X $ -direction, leaving the transverse directions $ x, y , p_x, 
p_y $
intact ;  
i.e. the  $ U ( 1, 1 ) =  SU ( 1, 1 ) \otimes U ( 1 )  $ subgroup
transformations 
that  leave invariant the  interval in a  classical ( commuting)  Phase
space  
are ( in units of $ \hbar = c = 1 $  )  :

$$  (d Z )^2 =   ( dT)^2  - ( dX)^2  +   {   ( dE)^2 - ( d P )^2   
\over b^2
} =
 invariant   = $$ 
$$ ( d \tau )^2   [ 1   +  {   ( dE/ d \tau )^2 - ( d P/ d \tau  )^2
\over b^2 }  ]  =  
( d \tau )^2  [   1  -  {  m^2 g^2 ( \tau )  \over m_P^2  A^2_{max}    }
]   . ~~~ b^2 \equiv m_P^2 A^2_{max} \eqno ( 3.1 ) $$
where  $ m_P$ is the Planck mass  $ 1 / L_P$ so that $ b = ( 1 / L_P 
)^2 $ ,
where $ L_P$ is the Planck scale.
 The quantity $ g ( \tau ) $  is the  proper four-acceleration of a 
particle
of mass $ m $ in the
$ x_3$-direction  which we take to be $X$ .  We have used the results 
of
eqs- ( 2-8, 2-9, 2-10   )
with $ ( d \tau)^2 = ( dT )^2 - ( dX)^2 $.
It is crucial to notice that  the particle`s   mass $ m $ is no longer 
an
invariant Casimir .
Only ratios of masses have physical meaning in Scale Relativity  [ 2 ] 
.
The  invariant interval $   ( dZ)^2 $  in eq- ( 3-1 )  is not the 
$same$ as
the interval 
$ ( d \omega )^2 $  of the Nesterenko action eq- ( 2-10 )
which is invariant under a pseudo-complexification of the Lorentz group  
[ 8
] .  
Only when $ m = m_P$,  the intervals agree.  The interval $ ( dZ)^2 $
described by Low [ 14 ] is
$ U ( 1, 3 ) $-invariant  for the most general transformations in the $ 
8D$
phase-space.

The transformations laws of the coordinates  in classical phase space 
are  [
14 ]  :

$$ T ` =  T cosh \xi   +  (  \xi_v  X + {  \xi_a P  \over b^2 }  )  { 
sinh
\xi \over \xi } . \eqno ( 3.2a ) . $$.

$$ E ` =  E  cosh \xi   +  (  - \xi_a  X +   \xi_v  P   ) { sinh \xi 
\over
\xi } . \eqno ( 3.2b ) $$

$$ X ' =  X  cosh \xi  +  ( \xi_v  T   -  {  \xi_a  E  \over b^2 }  )  
{
sinh \xi \over \xi } . \eqno ( 3.2c ) $$

$$ P ` =  P  cosh \xi  +  ( \xi_v  E  +  \xi_a T    )   { sinh \xi 
\over \xi
}  . \eqno ( 3.2 d) $$

The $ \xi_v $ = velocity-boost rapidity parameter  and $ \xi_a$ =
acceleration-boost 
rapidity parameter  of the primed-reference frame are defined 
physically as
: 

$$   tanh { \xi_v \over c }  =  \pm { v \over c } .~~~ tanh { \xi_ a 
\over b
}   = 
\pm  { a \over A_{max} } . \eqno ( 3.3 )  $$

The $effective$ boost parameter $ \xi $  of the $ U(1,1)$ subgroup
transformations appearing
in eqs- ( 3-2 ) is defined  in units of $ \hbar = c = 1 $  as  :

$$ \xi  \equiv  \sqrt {  \xi_v^2 +  {  \xi_a^2 \over b^2 } }  . \eqno ( 
3.4
) $$   
Our definition of the rapidity parameters are $different$ than those in 
[ 14
] . 
Straightforward algebra allows us to verify that :

$ \bullet $ The  transformations  given by  eqs-(3-2) leave the 
interval  of
eq-( 3-1 )  in classical phase space invariant.

$ \bullet $  The transformations  eqs-(3-2)  are fully consistent with
Born's duality symmetry principle  [ 14 ]
$ ( Q, P )  \rightarrow  ( P,  - Q )  $  .  By inspection we can see 
that
under  
Born duality the  transformations  in eqs- ( 3-2 )  are  $rotated$ into 
each
other, 
up to  numerical  $ b $ factors in order to match units.

$\bullet$ When on sets $ \xi_a = 0 $  in eqs- ( 3-2 ) one recovers
automatically the standard Lorentz transformations for the $ X, T $  
and $
E, P $ variables $separately$,  leaving invariant the intervals
$  dT^2 - dX^2 = ( d \tau)^2  $ and $  ( dE^2 - dP^2) / b^2  $ 
separately .
Naturally their $sum$ will also be maintained  invariant.

$ \bullet$ When one sets $ \xi_v = 0 $ we obtain the transformations 
rules
of the events in Phase space,
from one reference-frame into another  $uniformly$-accelerated frame of
reference,  
$ a = constant $ ,  whose  acceleration-rapidity parameter is  in this
particular case : 

$$ \xi  \equiv  { \xi_a \over b } . ~~~ tanh \xi  = {  a  \over A_{max} 
} .
\eqno ( 3.5 ) $$   
 
The transformations  for pure acceleration-boosts in one-spatial 
dimension
are  then :

$$ T ` =  T cosh \xi   + {  P \over b }  sinh \xi .  \eqno ( 3.6 a )  
$$.

$$ E ` =  E  cosh \xi   -  b X  sinh \xi .  \eqno ( 3.6 b )  $$

$$ X ` =  X  cosh \xi   -  {  E  \over b  }  sinh \xi .    \eqno ( 3.6 
c )
$$

$$ P ` =  P  cosh \xi  +   b T    sinh \xi   .  \eqno ( 3.6 d ) $$

It is straightforwad to verify that the transformations in eqs- ( 3.6 )
leave invariant the phase space
interval ( 3-1  ) but $do ~ not $ leave  $separately$ invariant the 
proper
time interval 
$  ( d \tau )^2 = dT^2 - dX^2 $.  Only the  $combination$  :

$$  ( d  Z )^2 =   ( d \tau )^2  (  1  -  {  m^2  g^2 \over  m_P^2
A^2_{max} }  )  . \eqno ( 3.6 e )  $$
is truly  left invariant under pure acceleration-boosts ! .

One can verify as well that the transformations ( 3-7 )  satisfy  
Born's
duality  symmetry  

$$  ( T, X ) \rightarrow  ( E, P )  . ~~~   ( E, P ) \rightarrow  ( -  
T, -
X ) .  \eqno ( 3.7 )  $$
and   $  b \rightarrow  { 1 \over b } $ . The latter duality 
transformation
is  nothing but a manifestation of a  large/small Tension duality 
principle
!  reminiscent of  the $ T$-duality in string theory;  i.e.  namely ,
small/large radius  duality = winding modes/ Kaluza-Klein modes duality 
in
string compactifications  and Ultraviolet/Infrared entanglement in
Noncommutative Field Theories.
Hence, Born's duality principle  in exchanging coordinates for momenta 
could
be the underlying physical reason behind $ T$-duality in string theory.

The Group property  of eqs-(3-6) is satisfied  i.e. the composition of 
two
succesive pure acceleration-boosts is another pure acceleration-boost 
with
acceleration rapidity given by $ \xi ``  = \xi + \xi ` $.  The addition 
of
$proper$ accelerations follows the usual relativistic composition rule 
:

$$  tanh \xi ``  = tanh ( \xi + \xi ` ) =  {  tanh \xi + \tanh \xi ` 
\over
1 + tanh \xi  tanh \xi ` }
\Rightarrow   {  a  ``  \over A }  =   {  {  a  \over A }  + { a `  
\over A
}  \over  1 +  { a  a `  \over A^2 }   } . \eqno ( 3.8 ) $$
in this fashion the upper limiting $proper$ acceleration is never
$surpassed$ like it happens with the ordinary Special Relativistic 
addition
of velocities. 

The  group properties of the  full  combination of velocity $and$
acceleration boosts ( 3-2) requires  much more algebra .    A careful 
study
reveals that the composition  $rule$   of two succesive transformations
given by eqs-(3-2 ) is  $preserved$  if,  and only  if ,  the following
four relationships among the
 $\xi ; \xi ` ; \xi  `` , ...... $   parameters are obeyed  :

$$ cosh \xi ``  =  cosh \xi cosh \xi `  + {  [  \xi_v \xi_v ` +    (  
\xi_a
\xi_a ` /  b^2 ) ]   \over   \xi \xi ` }  sinh \xi sinh \xi `  .    
\eqno (
3.9 a ) $$

$$ { \xi_v `` \over \xi ``  } sinh \xi ``  =   {  \xi_v  cosh \xi `   
sinh
\xi \over \xi } +  
{  \xi_v `   cosh \xi   sinh \xi `  \over \xi ` } . \eqno ( 3.9 b ) $$

$$ { \xi_a `` \over  b \xi ``  } sinh \xi ``  =  {  \xi_a   cosh \xi `
sinh \xi \over  b \xi } +
 {  \xi_a `  cosh \xi   sinh \xi  `   \over  b \xi `  }  . \eqno ( 3.9 
c )
$$

$$  \xi_a `  \xi_v - \xi_a \xi_v ` = 0 \Rightarrow   { \xi_a ` \over 
\xi_a }
= { \xi_v ` \over \xi_v } = \lambda .
\eqno ( 3.9 d )    $$

The condition of eq- ( 3.9 d )  can be recast as a global scaling of 
the
effective boost parameters as follows :
$$ \xi ` =  { \xi `  \over \xi } \xi = \lambda  \xi. ~~~ \lambda \equiv 
{
\xi ` \over \xi }  \Rightarrow
 { \xi_a `  \over \xi_a }  = { \xi_v `  \over \xi_v } =    { \xi  `   
\over
\xi }   =    \lambda .  \eqno ( 3.10 ) $$
Meaning that the primed variables are all rescaled by the same factor 
of $
\lambda $. 
From ( 3-10) we can infer :

$$ { \xi_a `   \over  \xi ` }  =  { \xi_a  \over  \xi } \Rightarrow  {
\xi_a `  \xi_a  \over b^2  \xi \xi `  }  =
(  { \xi_a \over b \xi }  )^2 .    \eqno ( 3.11 a )  $$

 $$ { \xi_v `  \over  \xi ` }  =  { \xi_v  \over \xi } \Rightarrow  {  
\xi_v
`   \xi_v  \over   \xi \xi ` }  =
(  { \xi_v  \over   \xi }  )^2 .  \eqno ( 3.11b  ) $$

As a result of eqs- ( 3-11a, 3-11b  ),  the definition of the effective
boost parameter given in eq- ( 3-4 ),  and
after using eq- ( 3-9a ),  we get :

$$ cosh \xi  ` `  =  cosh \xi cosh \xi  `  + { [   \xi_v \xi_v `  +  ( 
\xi_a
\xi_a ` / b^2 )  ]  \over   \xi \xi ` }
sinh \xi sinh \xi `   = $$
$$ cosh \xi cosh \xi `  +  1  sinh \xi sinh \xi `   =  cosh ( \xi +  
\xi ` )
\eqno ( 3.12a ) $$
Hence, as expected , we have found that the $effective$  boost 
parameters
are indeed $additive$  :

$$ \xi ``  = \xi +  \xi ``    \Rightarrow  sinh ( \xi  ``  ) = sinh ( 
\xi +
\xi ` ) = 
sinh \xi cosh \xi ` + cosh \xi sinh \xi `    \eqno ( 3.13 ) $$

From eqs- ( 3-9, 3-10, 3-11, 3-13 ) we can deduce :

$$ { \xi_a `   \over  \xi ` }  =  { \xi_a   \over \xi }   =   {   \xi_a 
``
\over \xi  ``   }   =  { \xi_a ``  \over  \xi + \xi ` }  =
{ \xi_a  ``   \over   \xi  ( 1 + \lambda )  } . \eqno ( 3-14 a ) $$

$$ { \xi_v  `   \over  \xi `  }  =  { \xi_v  \over \xi }   =     {   
\xi_v
` `   \over \xi  ` `    }   =  { \xi_v  ` `  \over  \xi + \xi ` }  =
{ \xi_v  ` `  \over    \xi  ( 1 + \lambda )  }. \eqno ( 3.14 b )  $$

Finally we arrive at the explicit expressions  for $ \xi ` `  ;  \xi_v  
` `
;  \xi_a  ` ` $ 
in terms of  the  other parameters

$$  \xi_v  ` `  =  \xi_v +  \xi_v  `  =    ( 1 + \lambda ) \xi_v. ~~~ 
\xi_a
` `  =  
\xi_a + \xi_a `  =  ( 1 + \lambda ) \xi_a  .
 \eqno ( 3.15 a )  $$

$$  \xi ` `  = \xi + \xi `   = ( 1 + \lambda ) \xi .  \eqno ( 3.15b  ) 
$$

This is all we  need to iterate again the group transformation rules to 
show
that  the composition of two succesive transformations with parameters 
$ \xi
` ;  \xi  ` `   , ...  $ yields another transformation with
parameters  given by $ \xi  ` ` ` = \xi `  + \xi  ` `   ; .....$.

Concluding, the $ Group$ law of the transformations eqs- ( 3-2 ) has 
been
explicitly proven. 
Hence, we truly have a Maximal-Acceleration Phase Space Relativity 
theory.
In the next section we shall see why this Relativity theory
can be translated as a Minimal Planck-Area Relativity Theory, for 
$minimal$
areas $L^2_P$ . 
And from Relativity in Phase spaces ( cotangent bundles of two-dim 
Riemann
surfaces, for example, )
we may have a new understanding of what
$ { \cal W } $ Geometry  ,  $ W_\infty$ strings,  $ { \cal W}$-gravity,
Higher conformal spin field theories in Anti de Sitter spaces, ....
may be telling us  [ 15, 16 , 17, 18 ] .   This is discussed next.
Once again, we emphasize that it is important to notice that our 
definitions
for the 
velocity/acceleration rapidity parameters  in eqs- ( 3-4, 3-5    ) are 
very
different from  those used by [ 14 ] .

\bigskip

\centerline{\bf  4. Planck-Scale Areas are Invariant  under  
Acceleration-
Boosts   } 

\bigskip 

Having displayed explicity the Group transformations rules of the
coordinates in Phase space
we will show why  $infinite$ acceleration-boosts ( which is $not$ the 
same
as infinite proper acceleration !  ) preserve Planck-Scale  $ Areas $
as a result of the fact that $ b = ( 1 / L_P^2 )  $ equals the $maximal 
~
invariant$  Force, or Tension,
if the units of $ \hbar = c = 1 $ are used.

At Planck scale $ L_P$ intervals we have by definition   (  in units of 
$
\hbar = c = 1 $ )  :

$$ \Delta X = \Delta T = L_P. ~~~ \Delta E = \Delta P = { 1\over L_P} . 
~~~
b \equiv  { 1 \over L_P^2 } = Maximal ~ Tension.  \eqno ( 4.1 )  $$

From eqs- (   3-6  )  we get after a direct use of  eq- ( 4-1 ) ,  in 
the
$infinite$  boost limit
$ \xi \rightarrow  \infty$,  :

$$ \Delta T `  =   L_P (  cosh \xi + sinh \xi )  \rightarrow \infty .
\eqno ( 4.2 a )  $$.

$$  \Delta E ` =   { 1 \over L_P } (   cosh \xi  - sinh \xi )  
\rightarrow 0
.   \eqno ( 4.2 b )  $$
by a simple use of L'Hopital's rule or  by noticing that both $ cosh 
\xi ;
sinh \xi  $ 
functions approach infinity  at the same rate.

$$ \Delta X ` =  L_P (  cosh \xi  - sinh \xi )  \rightarrow 0   .  
\eqno (
4.2 c )  $$

$$ \Delta P `  =  { 1 \over L_P }  (  cosh \xi  + sinh \xi )  
\rightarrow
\infty   . \eqno ( 4.2d )   $$
where  the discrete displacements of two events in Phase Space are 
defined
:

$$  \Delta X = X_2 -  X_1  = L_P . ~~~  \Delta E = E_2 - E_1 = { 1 
\over
L_P}  . $$ 
$$ \Delta T = T_2 - T_1 = L_P  ~~~ \Delta P = P_2 - P_1 = { 1 \over 
L_P}.
\eqno ( 4.3 ) .   $$

Due to the identity  :

$$  \infty \times 0 =  ( cosh \xi + sinh \xi ) ( cosh \xi - sinh \xi ) 
=
cosh^2 \xi - sinh^2 \xi = 1 . \eqno ( 4.4 ) $$
one can see from eqs- ( 4.2  ) that the Planck-scale $Areas$  are truly
$invariant $  
under the $infinite$ acceleration-boosts $ \xi = \infty$  :

$$ \Delta X ` \Delta P `   =  0 \times \infty =  \Delta X \Delta P  ( 
cosh^2
\xi - sinh^2 \xi )  =  \Delta X \Delta P =
{ L_P \over L_P } = 1 .
~~~ \hbar = c = 1 . \eqno ( 4.5 ) $$

$$ \Delta T `  \Delta E `   =  \infty \times 0 = \Delta T \Delta E    (
cosh^2 \xi - sinh^2 \xi )  = \Delta T \Delta E =
{ L_P \over L_P } = 1.
~~~ \hbar = c = 1. \eqno ( 4.6 )  $$

$$ \Delta X ` \Delta T  `   =  0 \times \infty =  \Delta X \Delta T  (
cosh^2 \xi - sinh^2 \xi )  =  \Delta X \Delta T  =
  ( L_P )^2  . \eqno ( 4.7 ) $$

$$ \Delta P `  \Delta E `   =  \infty \times 0 = \Delta P  \Delta E    
(
cosh^2 \xi - sinh^2 \xi )  = \Delta P  \Delta E =
{ 1 \over L_P^2  }  . \eqno ( 4.8 )  $$

It is crucial to emphasize that the invariance property of the minimal
Planck-scale $ Areas $
( maximal Tension ) is  $not$ an exclusive property of $infinite$
acceleration boosts $ \xi = \infty$,  but,  as a result of the identity   
$
cosh^2 \xi - sinh^2 \xi = 1  $ , for all values of $ \xi $ , the  
minimal
Planck-scale $ Areas$ are  $always$  invariant under  $any$
acceleration-boosts transformations  !. Meaning physically, in units of 
$
\hbar = c = 1 $ , that the Maximal Tension ( or maximal Force )  $ b = 
{ 1
\over L_P^2 } $ is a true physical $invariant$  universal quantity !  [ 
15
] .   Also we notice that the Phase-space areas,  or cells ,  in units 
of
$ \hbar$,  are also invariant !  The pure-acceleration boosts
transformations  are " symplectic " .

The infinite acceleration-boosts  are closely related to the
infinite red-shift effects  when light signals barely escape Black hole
Horizons  reaching an asymptotic observer with an infinite redshift 
factor.
The important fact is that the Planck-scale $Areas$ are truly 
maintained
invariant under acceleration-boosts. This  could reveal  very important
information about 
Black-holes  Entropy and Holography .    The logarithimic corrections 
to the
Black-Hole Area-Entropy relation were obtained directly from
Clifford-algebraic methods in C-spaces [ 1 ] , in addition to the 
derivation
of the maximal Planck temperature condition and the Schwarzchild radius 
in
terms of the Thermodynamics  of a gas of p-loop-oscillators  quanta :
area-bits, volume-bits, ... hyper-volume-bits in Planck scale units.
  
\bigskip 

\centerline { \bf Concluding Remarks }

\bigskip 

To finalize we make some important remarks pertaining the Conformal 
group,
the physics of branes,
$W$ algebras , Higher conformal-spin field theories, ....within the 
context
of Relativity of Phase spaces and the group of minimal Planck-Area
Relativity. 

The conformal algebra $ SO(4, 2 ) $ in four-dimensions can be extracted
$directly$ from  the $ D = 4 $ Clifford algebra $without$  the need to 
recur
to $ 5-dim$ hyperboloids embedded in $ D = 6 $  [ 1 ]  .
The conformal algebra in $ D = 4 $ is isomorphic to the isometry 
algebra of
$ AdS_5$. 
The conformal group $ SO(4, 2 ) $ has $15$ generators like the group $ 
SU(1,
3 ) $ has. 
In fact,  pure acceleration boosts play a similar role as 
conformal-boosts (
special conformal transformations )
since uniformly-accelerated trajectories in flat space are given by
hyperbolas.   

The conformal  agebra in $ D = 2 $ is infinite-dimensional and further
extensions of the conformal algebra in $ D = 2 $  exist which
are given by $ W$ algebras. The latter are deeply ingrained with the 
algebra
of symplectic-diffs ( area-preserving ).  Higher conformal spins 
$W_\infty,
W_{1+ \infty} $ -algebras  (  in $ D = 2 $ )  [ 17, 18  ]
are associated with the area-preserving diffs of a plane and cylinder
respectively.  

The $ SU(\infty)$ algebras are the area-preserving diffs of sphere in a
suitable $basis~ dependent $ limit  [ 19 ] .  Yang-Mills theories in $ 
D = 4
$ are conformally  invariant and in [  21 ] we have shown why $ 
p$-brane
actions can be obtained from a Moyal deformation quantization of (
Generalized ) $SU(N)$ Yang-Mills theories, where the auxiliary phase 
space
variables required in the Moyal deformation procedure are later 
identified
with the world-volume coordinates of the $ p$-branes.   The large  $ N$
limit is equivalent to the
classical $ \hbar = 1/N \rightarrow 0 $ of the Moyal-bracket algebra.

Infinite-dimensional extensions of the $finite$-dim  conformal algebra 
in $
D > 2 $ exist and were constructed by Vasiliev et al  [ 15 ] .  The
construction of higher spin theories on $ AdS_D$ spaces  can be 
attained
also by a Noncommutative Moyal-like star product deformation of the
symplectic algebras  ( oscillator algebras ) in
$ D > 2 $ , whose deformation parameter is the inverse of the throat 
size of
$ AdS_D$ space.   

The fact that $ U(1,3 )$ is the symmetry group of classical phase space
Relativity [ 14 ],  and that the canonical group $ C ( 1, 3 ) $ 
contains the
$oscillator$  algebra in four dimensions is very appealing [ 14 ] .  It 
is
is also consistent with the fact that  the $ U ( 2, 2 ) $ 
tensor-operator
algebras of higher conformal-spin field theories on $ AdS_D$ spaces  
have
been proposed  by Calixto [ 16 ] as the $higher$-dimensional candidates 
of
$ W$-like algebras that are essential in the construction of  $induced$
conformal gravities in $higher$ dimensions.  This would be a 
generalization
of  the  WZNW ( Wess, Zumino, Novikov, Witten )  models  to higher
dimensions.   

The $W$-geometry of the cotangent bundle ( phase space ) of  two-dim (
complexified ) Riemann surfaces
 [  17  ]  was shown to be directly connected to the 
Fedosov-deformation
quantization  procedure  of symplectic manifolds [   20 ] ,  which is
required when the Phase spaces are $curved$ .
A $curved$ Phase space would be the extension of General Relativity in
ordinary spacetimes.   Instead of studying ordinary strings one may be
forced to look deeply into $ W_\infty$ strings moving in curved 
backgrounds
, like $ AdS$ spaces.  Noncritical $ W_\infty$  bosonic ( super ) 
strings
are 
effective $ 3D$ theories devoid of BRST anomalies in $ D = 27,  11  $
dimensions , respectively [ 20 ] ,
which coincide with the allegedly critical dimensions of the bosonic ( 
super
) membrane.  
Hence, noncritical $ W_\infty$ strings,  effective $ 3D$ theories,  
behave
like critical membranes [ 20  ] .
And moreover,  they live on the (  proyective ) $ 3D $  conformal
$boundary$  of $ AdS_4 $.

In view of all these interesting connections related to the algebras of
area-preserving diffs in Phase-Space ,
it is warranted to study the full C- Phase-Space Relativity Theory , 
and its
algebra , in order to construct a
unified theory of  $all$  p-branes.  We believe that C-space Extended 
Scale
Relativity Theory  is the  appropriate arena to study the physics of
$p$-loops = closed p-branes , for $ all$ values of $ p $  [ 1  ] .

Concluding,  it  seems that Quantum Gravity is deeply linked to the 
Geometry
of 
Noncommuting  Phase spaces rather than with the  naive Quantization of 
a
spacetime , 
and its metric , in spacetimes of $fixed$ dimension.  The dimensions 
and
signatures are also variables in
C-space Extended Relativity  Theory [ 1 ] , i.e   $all$ dimensions and
signatures are treated on equal footing.
Since the notion of a " point " is lost due to the minimal Planck 
length
principle,  the notion of fixed dimension and fixed Topology  is also
naturally lost.  Roughly speaking,  instead of points  living a 
particular
space  of fixed dimension ,  we have  " Dimensional"  and " 
Topological"
fluctuations/oscillations  within all the p-loop oscillations of 
C-space.

\bigskip 

\centerline{\bf  Acknowledgements}

We are kindly indebted to M.Bowers ,  H. Rosu and J. Mahecha   for 
their
assistance in
preparing the manuscript and hospitality in Santa Barbara where this 
work
was completed.  
We thank S. Low for sending us reference [ 14 ] .

\bigskip

\centerline { \bf References }

\bigskip

1 - C. Castro, ``The programs of the Extended Relativity in C-spaces,
towards
the physical foundations of String theory", hep-th/0205065. To appear 
in the
proceedings of the NATO advanced workshop on the nature of time, 
geometry,
physics and perception. Tatranska Lomnica, Slovakia, May 2002.  Kluwer
Academic Publishers.  Noncommutative Quantum Mechanics and Geometry 
from the
quantization of C-
spaces", hep-th/0206181.  ³ Maximal-Acceleration Phase Space Relativity 
from
Clifford Algebras ³ hep-th/0208138 .  " Variable Fine structure 
constant
from Maximal-Acceleration Phase space Relativity " hep-th/0210 
\noindent

C. Castro, M. Pavsic : ³ Higher Derivative Gravity and Torsion from the
Geometry of C-spaces ³
Phys. Lett { \bf  B 539 } ( 2002 ) 133.  hep-th/0110079.   ³ The 
Clifford
Algebra of spacetime and the conformal group ³ hep-th/0203194.

2 - L. Nottale, ``Fractal Spacetime and Microphysics, towards Scale
Relativity", World Scientific, Singapore, 1992; ``La Relativite dans 
tous
ses etats", Hachette Literature Paris, 1999.

3 - M. Pavsic, ``The landscape of Theoretical Physics: A global view  
from
point particles to the brane world and beyond, in search of a unifying
principle",  Kluwer Academic Publishers 119, 2001.

4 - E. Caianiello, ``Is there a maximal acceleration?", Lett. Nuovo 
Cimento
{\bf 32} (1981) 65.

5 - V. Nesterenko, Class. Quant. Grav. {\bf 9} (1992) 1101; Phys. Lett. 
{\bf
B 327} (1994) 50; V. Nesterenko,  A. Feoli, G. Scarpetta, ``Dynamics of
relativistic particle "

6-H. Brandt :  Contemporary Mathematics {\bf 196 } ( 1996 ) 273.
Chaos, Solitons and Fractals {\bf 10 } ( 2-3 ) ( 1999 ) .

7-M.Pavsic : Phys. Lett {\bf B 205 } ( 1988) 231 ; Phys. Lett {\bf B 
221} (
1989 ) 264.
H. Arodz , A. Sitarz, P. Wegrzyn : Acta Physica Polonica { \bf B 20} ( 
1989
) 921.

8- F. Schuller :   `` Born-Infeld Kinematics and corrections to the 
Thomas
precession ''
hep-th/0207047, Annals of Phys. {\bf 299 } ( 2002 ) 174.

9- A. Chernitskii : `` Born-Infeld electrodynamics, Clifford numbers and
spinor representations '' hep-th/0009121.

10- J. Lukierski, A. Nowicki, H. Ruegg, V. Tolstoy : Phys. Lett { \bf 
264 }
( 1991 ) 331. 
J. Lukierski, H. Ruegg , W. Zakrzewski : Ann. Phys { \bf 243 } { 1995 ) 
90.

G.Amelino-Camelia :  Phys. Lett { \bf B 510} ( 2001 ) 255.
Int. J. Mod. Phys { \bf D 11 } ( 2002 ) 35, gr-qc/0012051

11- S. Rama : `` Classical velocity in kappa-deformed Poincare algebra 
and a
Maximal Acceleration '' hep-th/0209129.

12- J.P. Uzan, `` The fundamental constants and their variations :
observational status and
theoretical motivations '' hep-ph/0205340 .

13 - G. Lambiase, G.Papini. G. Scarpetta :  `` Maximal Acceleration
Corrections to the Lamb Shift of one Electron Atoms '' hep-th/9702130.

14- M. Born  : Proc. Royal Society {\bf A 165 } ( 1938 ) 291; Rev. Mod.
Physics { \bf 21 } ( 1949 ) 463.
 
S. Low :  Jour. Phys { \bf A  } Math. Gen {\bf 35 }  ( 2002 ) 5711.
\noindent 

C. Kalman :  Can. Jour. Physics { \bf 51 } ( 1973 ) 1573.

15- M. Vasiliev : " Higher Spin Gauge theories , Star Products and AdS
spaces " hep-th/9910096.

16- M. Calixto : " Higher $ U(2, 2 ) $ spin fields and higher-dim
$W$-gravities .... " hep-th/0102111.

E. Nissimov, S. Pacheva , I. Vaysburd :  " $W_\infty$ Gravity,  a 
geometric
approach " hep-th/9207048.

17. C. Hull : Phys. Letts {\bf B 269 } ( 1991 ) 257.

18.  P. Bouwkgnet , K. Scouetens : "  $ W$ symmetry in Conformal Field
Theory " 
Phys. Reports {\bf 223 } ( 1993 )  183-276.

E. Segin :  " Aspects of $W_\infty$ symmetry, hep-th/9112025.

19. J. Hoppe :  " Quantum theory of a Relativistic Surface : MIT , Ph. 
D
thesis 1982.
\noindent

20- C. Castro ,  " W-geometry from Fedosov deformation quantization : 
J.
Geometry and Physics
{\bf 33 } (  2000 ) 173.   J. Chaos, Solitons and Fractals {\bf 7 } ( 
1996 )
711.   
" A New Realization of Holography  " hep-th/0207231.

21- C. Castro : " Branes from Moyal deformation quantization of 
Generalized
Yang-Mills " hep-th/9908115.

S. Ansoldi, C. Castro, E. Spallucci :  Class. Quant. Grav { \bf 18 } ( 
2001 )  2865.

\end{document}